\documentclass{aa_new}

\usepackage{graphicx}

\newcommand{\vd}{\overrightarrow{d}}

\newcommand{\vTeta}{\overrightarrow{\Theta}}
\newcommand{\vTetab}{\overrightarrow{\Theta}_{best}}
\newcommand{\mC}{\vec{C}}

\newcommand{\Npix}{N_{\rm pix}} 
 
\newcommand{\Nband}{N_{\rm band}} 
\newcommand{\fbP}{\delta T_{\rm fb}}

\newcommand{\cf}{characteristic function } 
 
\newcommand{\dtt}{\delta T^2}

\def\be{\begin{equation}} 
\def\ee{\end{equation}} 
\def\bea{\begin{eqnarray}} 
\def\eea{\end{eqnarray}}         
\def\bc{\begin{center}} 
\def\ec{\end{center}}

\begin{document} 
 
%
   \title{Goodness--of--fit Statistics and CMB Data Sets} 
%
 
        \titlerunning{Goodness of fit in  CMB}  
 
   \author{ M.~Douspis$^{1,2}$, J.G.~Bartlett$^{1,3}$, A.~Blanchard$^{1}$}

   \offprints{douspis@astro.ox.ac.uk} 
 
   \institute{$^1$ Observatoire Midi-Pyr\'en\'ees, 
              14, ave. E. Belin, 
              31400 Toulouse, FRANCE \\ 
              Unit\'e associ\'ee au CNRS  
             ({\tt http://webast.ast.obs-mip.fr})\\     
              $2$ Astrophysics, 
                Nuclear and Astrophysics Laboratory 
                Keble Road 
                Oxford OX1 3RH, 
                UNITED KINGDOM\\ ({\tt http://www-astro.physics.ox.ac.uk})\\
		$2$ APC, Universit\'e Paris 7, Paris
                FRANCE 
                ({\tt http://apc-p7.org/})
		             } 
 
  \date{Submitted: June 2001, Accepted May 2003}

   \abstract{ 
         Application of a {\em Goodness--of--fit} (GOF) statistic 
is an essential element of parameter estimation.  
We discuss the computation of GOF when estimating parameters from  
anisotropy measurements of the cosmic microwave background (CMB), 
and we propose two GOF statistics to be used  
when employing approximate band--power likelihood functions. 
They are based on an approximate form for the distribution 
of band--power estimators that requires only minimal experimental 
information to construct.  Monte Carlo simulations of 
CMB experiments show that the proposed form describes  
the true distributions quite well.  
We apply these GOF statistics to current CMB anisotropy data  
and discuss the results.
} 
\maketitle 
 
\section{Introduction} 

     Measurement of the cosmic microwave background (CMB) temperature
anisotropies has proven to be one of the most powerful tools
for estimating important cosmological parameters (\cite{boom2}, \cite{pryke}, 
\cite{rubino}, \cite{sievers},
\cite{wang}).  The observed angular power spectrum shows the coherent 
peak structure expected in inflationary models, and fitting model curves
to the data\footnote{See {\tt http://webast.ast.obs-mip.fr/cosmo/CMB} 
for an up--to--date compilation.} yields constraints on many 
parameters.  This leads in particular to the conclusion that the 
geometry of space is flat (\cite{lineweaver}, \cite{boom},
\cite{maxima}, \cite{lange}, \cite{balbi}).  In terms of statistics,
the procedure just described is one of parameter estimation.  
 
     Parameter estimation proceeds via the identification 
of a {\em best} model (set of parameters) within a  
family of models, an evaluation of the quality of the fit 
and the construction of parameter constraints.  
The method of maximum likelihood (ML), for example, is a useful,  
general procedure for finding a best--fit model.
As a general rule, one must judge the quality of the fit 
before any serious consideration of parameter constraints. 
This requires the application of a {\em Goodness--of--fit} 
(GOF) statistic.  Such a statistic is, usually, some scalar function
of the data whose distribution may be calculated once given
an underlying physical model and a model of the statistical
fluctuations in the data.  
It is generally a function $gof(d,T)$ of both the data $d$ and theory
$T$, such that $gof$ attains, for example, 
a minimum when $d$ is generated by the theory $T$. 
It is defined in a 'monotonic' way, in the sense that $gof$ 
becomes larger as $d$ gets 'further' from a
realization of $T$. The 'significance' may then be 
defined as the probability of obtaining $gof>gof_{\rm obs}$.
On this basis, it permits a quantitative evaluation of 
the quality of the best model's fit to the data: if the probability
of obtaining the observed value of the GOF statistic (from
the actual data set) is low (low significance), then
the model should be rejected.  Without such a statistic, one does 
not know if the best model is a good model, or simply 
the ``least bad'' of the family. 
 
     In this {\em paper}, we examine in some detail 
the issue of GOF when analysing anisotropy data 
on the cosmic microwave background.  
The vast majority of present
analyses of the power spectrum data 
do not include proper GOF evaluations. 
The problem is particularly complicated by the fact that  
approximate likelihood methods must be employed in order to 
process the large volume of data and to explore 
a significant part of parameter space.  These methods 
usually rely on power spectrum {\em estimates}, such as  
flat band--powers, extracted either from scan data, 
or from reconstructed sky maps.  Because the power 
is quadratic in the temperature fluctuations, it 
is clear that these estimates are not Gaussian  
distributed.  The traditional approach of $\chi^2$ minimisation  
incorrectly assumes that power estimates are Gaussian  
distributed, something that can lead to a bias in  
determining the best model (e.g., Douspis et al. 2001a). 
For the same reason, the value of the reduced $\chi^2$  
at the best model does not retain its usual statistical  
meaning and may therefore not be simply used as a GOF statistic. 
 
     Approximations to the band--power likelihood function 
that permit more rigorous analyses have been proposed 
(Bond, Jaffe \& Knox 2000; Bartlett et al. 2001).  The question 
remains, however, of how to correctly evaluate the 
GOF of the best model.  Such an evaluation requires 
knowledge of the distribution of the power estimates, 
which is not necessarily the same as the likelihood 
function.  Using the same approach as Bartlett et al. (2001; 
hereafter paper 1), we propose an ansatz for the  
distribution of band--power estimates and  
test it against Monte Carlo simulations of certain 
MAX and Saskatoon data sets.  The ansatz requires only 
minimal experimental information, and it appears 
to work well.  We therefore use it to construct two  
GOF statistics, which we then apply to various  
ensembles of the present CMB data set.  
 
 
 
\section{Likelihood Method} 
 
     It is useful to begin with a discussion of GOF in 
the context of a complete likelihood analysis.  Although 
computationally challenging (in fact, impossible for 
large data sets: Bond et al. 2000; Borrill 1999ab),  
a likelihood approach is conceptually 
straightforward and our discussion serves to highlight 
certain important points.  Such an analysis is 
in any case required for a small subset of data in 
order to test approximate methods (see, for example, 
Douspis et al. 2001a, hereafter paper 2). 
 
 
     Following the notation of papers 1 and 2, we write the  
likelihood function as (we consider only Gaussian 
perturbations) 
\be 
\label{eq:like} 
{\cal L}(\vTeta) \equiv {\rm Prob}(\vd|\vTeta) 
        = \frac{1}{(2\pi)^{\Npix/2} |\mC|^{1/2}} e^{-\frac{1}{2} 
        \vd^t \cdot \mC^{-1} \cdot \vd}       
\ee 
where $\mC(\vTeta)$ is the correlation matrix (a function of  
the model parameters $\vTeta$ and including a contribution from 
instrumental noise), 
and  $\vd$ is column vector listing the pixel values\footnote{These 
`pixels' may either be the simple pixels of a map, or temperature 
differences, as given by, for example, MAX.}.  
The elements of $\vTeta$ may be either the cosmological parameters, 
or a set of band--powers. 
Maximising the likelihood function over the parameters  
defines the ``best model'' corresponding to  
the parameters $\vTetab$.  
 
     In the present situation, we are greatly aided by the 
Gaussian form of Eq. (\ref{eq:like}) {\em in the data vector}, $\vd$. 
Given the best model, the most obvious GOF statistic 
is then clearly  
\bc 
\be 
\label{eq:likechi2} 
gof = \vd^t \cdot \tilde{\mC}^{-1} \cdot \vd 
\ee 
\ec 
where $\tilde{\mC}\equiv C(\vTetab)$ is the correlation matrix evaluated  
at the best model.  For the Gaussian fluctuations we have assumed, 
this quantity follows a $\chi^2$ distribution, with a number 
of degrees--of--freedom (DOF)  approximately equal to the number of  
pixels minus the number of parameters\footnote{This 
recipe does not strictly apply in the present case, because 
the parameters are {\em non--linear} functions of the 
data; it is nevertheless standard practice.  In any case, 
the number of pixels is in practice much larger than the 
number of parameters.}. 


\begin{figure} 
\begin{center} 
\resizebox{\hsize}{!}{\includegraphics[angle=0,totalheight=9cm, 
        width=8.9cm]{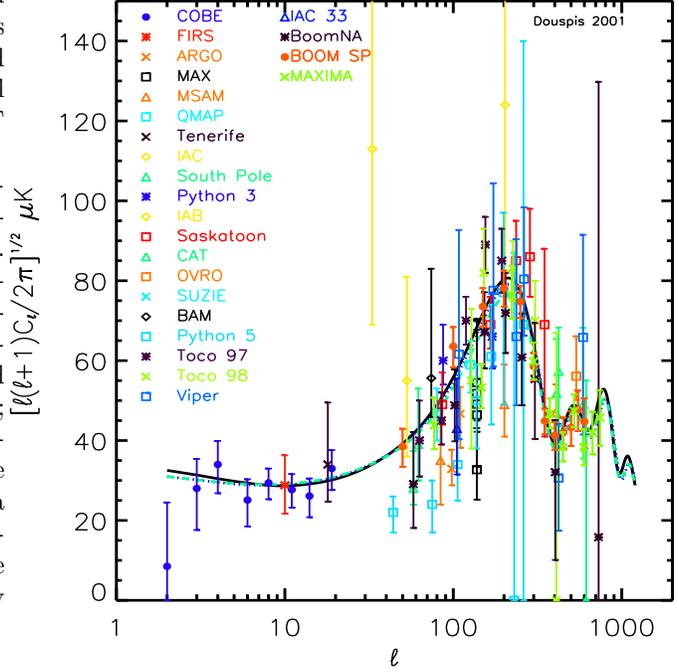}} 
\end{center} 
\caption{Power spectrum plot of some actual CMB data} 
\label{fig_plot} 
\end{figure}

\section{$\chi^2$ method} 
 
     For a variety of reasons (e.g., increased computational  
speed or inaccessible pixel data) most parameter estimations 
use power estimates, $\delta T^2$,  
as their starting point, such as those shown in  
Figure \ref{fig_plot}.   
A classic minimisation of $\chi^2$ 
\be 
\label{eq:classicchi2} 
\chi^2 (\vTeta)= \sum_{n=1}^{N_{exp}} \left( \frac{\delta T_{n}^{obs} 
    -\delta T_n(\vTeta)}{\sigma_n} \right)^2 
\ee 
is commonly used to find $\vTetab$ and the best model, 
where $\sigma_n=\sigma_+$ ($\sigma_-$) if the model 
passes above (below) the data point.  The obvious 
GOF statistic would then be the value of the $\chi^2$ 
evaluated at the minimum: $gof = \chi^2(\vTetab)$. 
As already noted, this whole procedure is inappropriate because 
power estimates do not follow a Gaussian distribution.  
It is of course true that if the number of 
contributing effective degrees--of--freedom \footnote{less than the number of  
original pixels by a factor depending on the pixel--pixel 
correlations; see paper 1} 
is large, a power estimate will closely follow a Gaussian; this, 
however, is never the case on the largest scales probed by 
a survey.  We shall see in the following that, for actual 
CMB data, the $\chi^2$ approach leads to quantitatively 
different results than other, more appropriate GOF statistics. 
For future reference, we show the 
value of this classic $\chi^2$ in Table 1.   
 
\section{Proposed approximation} 
 
     To improve  the $\chi^2$ analysis, several authors 
have proposed approximations to the band--power likelihood 
function ${\cal L}(\delta T)$ that may be constructed 
based on only minimal information about the experimental 
set--up (Bond, Jaffe \& Knox 2000; Paper 1).  One then 
arrives at the likelihood as a function of cosmological 
parameters $\vTeta$ with ${\cal L}[\delta T(\vTeta)]$.   
Unfortunately, these approximate likelihood functions 
do not retain the normalisation of the full likelihood 
over pixels (Eq.\ref{eq:like}).  This is a crucial point  
for GOF: we cannot deduce the quantity in Eq. (\ref{eq:likechi2}) 
from the value of the approximate  
likelihood at its maximum. 
 
     An alternative way to build a GOF statistic would 
be from the expected  
distribution of power estimates, i.e., the distribution 
of points in Figure 4 around the model curve.  Testing 
the observed dispersion of actual power points around 
the best--fit model against this expectation amounts 
to a GOF.  The main difficulty in this approach is that 
we do not have an expression for the distribution 
of ML power estimates.  It is important to understand 
that this distribution is not the same as the band--power 
likelihood, whose maximum is used to find the estimated power. 
In this section, we first motivate and then test an 
approximation to the distribution of ML power estimates. 
 
\subsection{Motivating an ansatz} 
 
     Our approach will be the same as in Paper 1, and 
the following results thus apply when using the approximate 
band--power likelihood introduced therein.  We motivated 
our likelihood approximation with an  
unrealistically simplified situation of $\Npix$  
uncorrelated pixels and uniform noise (refered to hereafter 
as the {\em simple picture}).  This 
suggested a functional form depending on two 
parameters, an effective number of degrees--of--freedom
$\nu$ and a noise parameter $\beta$; in the simple 
picture, $\nu=\Npix$ and $\beta^2$ is the noise variance.   
These two parameters could be found in realistic  
situations by adjusting to published flat--band 
confidence intervals (``errors'').  The particular 
advantage of such a technique  is that it permits an 
approximate likelihood analysis based on rather rudimentary 
information often found in the literature; this is an 
important advantage for many first generation experiments.   
In this same spirit, we 
now propose an ansatz for the ML band--power  
estimators. 
 
     For the simple picture ($\nu=\Npix$ and  
$\beta^2=$ noise variance), we showed in Paper 1 that the  
ML band--power estimator, $\delta T^2$, was a linear transform 
of a $\chi_{\Npix}^2$ random variable: 
\be 
\chi^2_\nu = \nu  \frac{([\delta T]^2 + \beta^2)} 
{([\delta T(\vTeta)]^2 + \beta^2)} 
\ee 
where $\delta T^2(\vTeta)$ is the band--power of the  
underlying model.  
{\em In a realistic situation where $\nu$ and $\beta$ are 
found from published power estimates, there is no 
a priori guarantee that this formula applies with  
the same values of $\nu$ and $\beta$}.   
One is, of course, tempted to suppose 
that the same values may in fact be used, at least approximatively. 
This hope forms the basis of our proposed ansatz  
for the band--power estimator distribution: 
\begin{eqnarray}\label{eq:approx} 
{\cal P}(\delta T^2|\vTeta) & \propto & Y^{(\nu/2 - 1)} e^{-Y/2} \\ 
\nonumber 
\\ 
\nonumber 
Y[\delta T^2] & \equiv & \nu . \frac{([\delta T]^2 + \beta^2)}{([\delta T(\vTeta)]^2 + \beta^2)}  \\ 
\nonumber 
\end{eqnarray}  
The underlying model band--power $\delta T^2(\vTeta)$ is  
in practice taken to be the ML estimate. 
The essential spirit of our approach is that, knowing the 
flat--band estimates and the 68 and 95\% confidence levels,  
one is able to reconstruct the entire 
likelihood function {\em and} (now) the probability  
distribution of the {\em estimate} $\fbP$.   
 
     The only way to be sure that this proposed method  
actually works is by testing it against Monte Carlo 
simulations of some experiments before generalised it.
  We mention at least 
one reason for caution: the quantity $\nu$ represents 
an effective number of DOF, reduced 
from $\Npix$ by inter--pixel correlations, applicable 
to the likelihood function; it is not at all clear that  
this same effective DOF applies equally well to the power  
estimator distribution (as it does in the simple picture). 
In particular, note that since the same data where 
used to find the best--fit model, we might expect  
 a reduction in DOF, something familiar 
from the classic reduced $\chi^2$ test.  Here, however, 
we have no clear idea of the reduction.   
Fortunately, the proposed method nevertheless appears valid, as 
the following Monte Carlo simulations demonstrate it. 
\begin{figure} 
\begin{center} 
\resizebox{\hsize}{!}{\includegraphics[angle=0,totalheight=9cm, 
        width=8.9cm]{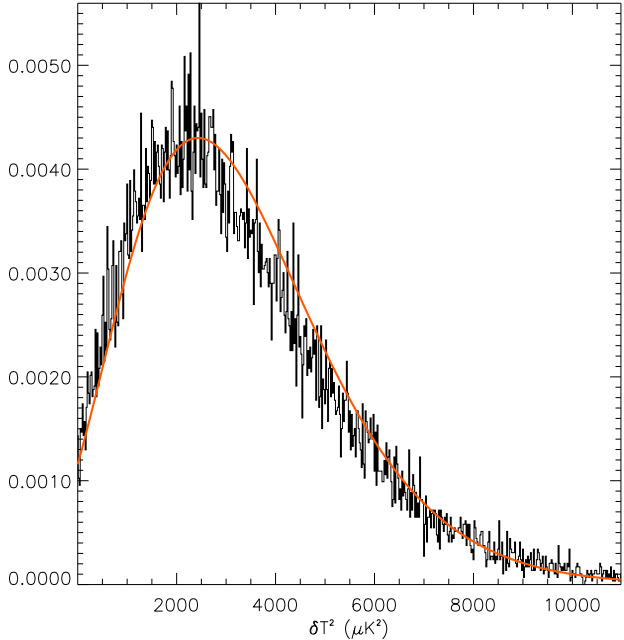}} 
\end{center} 
\caption{Distribution of the ML flat band--power  
estimator for $MAX\ ID\ 3.5\  
cm^{-1}$ found by Monte Carlo simulation.  The smooth 
(red) curve is the approximation Eq.~(\ref{eq:approx}), which 
fits the distribution well.} 
\label{fig_prob} 
\end{figure}

\subsection{Testing the ansatz} 
 
     We simulated many different data realizations of the 
MAX ID (Clapp et al. 1994) and Saskatoon (Netterfield et al. 1996) 
 experiments  
in order to reconstruct the corresponding ML power estimator distribution.
For example, we ran 30000 realizations of MAX ID 
at a frequency of $3.5 {\rm cm}^{-1}$ in the following 
manner:  we first compute the flat band--power and 
the one dimensional likelihood function for the actual 
observational data.  Knowledge of the latter provided  
the value of the pair $(\nu, \beta)$.  
The maximum of the likelihood function gave us the  
``best model'', which was used to simulate pixels  
on the sky. In order to take 
into account all correlations, we simulated our 
pixels using the full pixel--pixel correlation matrix.  
We first computed the theoretical part of the correlation  
matrix evaluated for our ``best model''.  After  
diagonalization, we drew 30000 realisations of  
21 pseudo--pixels from a Gaussian distribution  
centered on 0 and with the variances given by the eigenvalues.  
We reconstructed the ``true'' sky pixels using the transformation matrix  
(eigenvector matrix) and adding realizations of Gaussian noise  
(given by the known noise correlation matrix).  We thus obtained 
30000 sets of 21 pixels, correlated and drawn according to 
the best model ($\fbP=57.3 \mu K$).  For each realization, we 
derived the ML power estimate and build a histogram of its 
distribution. 
 
     Figure \ref{fig_prob} shows the resulting distribution for  
$MAX\ ID\ 3.5\ cm^{-1}$.  Overplotted in red as the 
smooth curve is the ansatz Eq. (\ref{eq:approx}) with the  
same values of $(\nu, \beta)$ as found from the likelihood  
function.  We see that the proposed approximate distribution  
is indeed a good representation of the true power estimator distribution. 
 
     The same kind of analysis was performed for the Saskatoon K--band  
3--point data, an altogether different observing strategy.  Once 
again, the approximation fitted the distribution to high accuracy. 
On the basis of these agreements, we will now adopt the proposed form  
in Eq. (\ref{eq:approx}) as a good representation of the  
distribution of ML band--power estimators.

\subsection{From probability function to GOF} 
 
     On the basis of the distribution Eq. (\ref{eq:approx}), 
we now construct two GOF statistics.  The goal is 
to define a scalar quantity, $gof$, that measures the scatter 
of points around a given model and whose distribution 
is known under the hypothesis that this model 
represents the ``truth'' (the null hypothesis). 
An improbable value of $gof$ would indicate that 
there is a problem. 
 
     Both constructions assume that the band--powers are 
independent.  This of course is not strictly true, 
but generally speaking published band--powers do not 
have strong statistical correlations; for example,  
the residual correlation between the Saskatoon bands is at 
a level of $\sim 10\% $.  Calibrations errors, on  
the other hand, do induce important band--band correlations. 
As already mentioned, the present work does not include 
calibration errors, and any ``bad fit'' indicated by  
our GOF tests could indicate either a false model, or 
that calibration errors are important.   
Our aim here is to show the ability of a proper GOF to 
identify problems with CMB power data fits, and to 
demonstrate the advantage of  the two proposed 
GOF statistics on the naive and inappropriate classic  
$\chi^2$. 

\subsubsection{Generalized $\chi^2$} 
 
   For each band--power $i$, consider the variables $\alpha_i$  
defined as follows: 
\be 
\int_{-\infty}^{\alpha_i} \frac{1}{\sqrt{\pi}}\exp(- x^2/2) dx = p_i 
\ee 
where $p_i\equiv \int_{-\beta_i^2}^{\delta T_i^2} {\cal P}_i(\dtt|\vTeta) 
d\dtt$ is calculated using Eq. (\ref{eq:approx}). 
The $\alpha_i$ are thus Gaussian random variables with zero mean and 
a variance of unity.  Hence, the sum $gof = \sum_1^{N_{exp}} \alpha_i^2$  
follows a $\chi^2$ distribution with $\Nband$  
DOF and provides a handy GOF statistic.   
 
\begin{figure}\label{fig_chi2} 
\begin{center} 
\resizebox{\hsize}{!}{\includegraphics[angle=0,totalheight=9cm, 
        width=8.9cm]{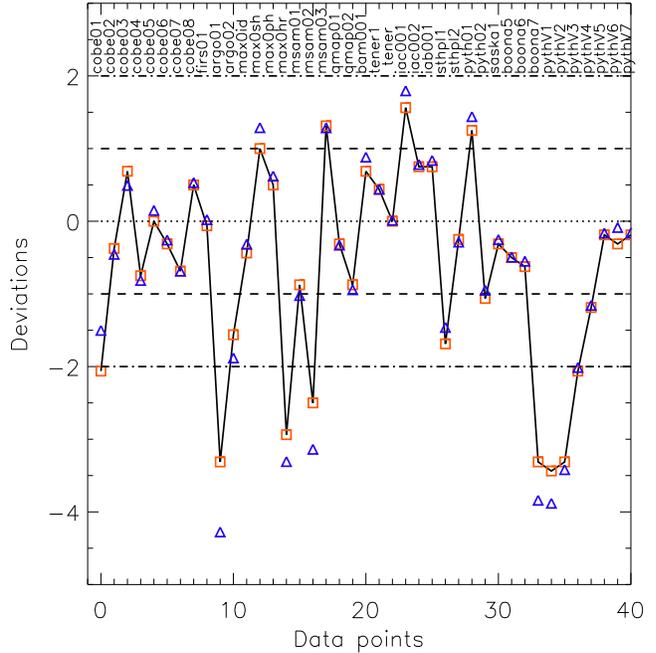}} 
\end{center} 
\caption{Individual $\nu_i$ (red squares) and $\chi^2$ (blue triangles) for a subset of the data plotted in figure 1.} 
\end{figure}

\subsubsection{Characteristic functions} 
 
     Another way to define a GOF statistic for a 
fit to $\Nband$ power points relies on the following property  
of characteristic functions: {\it the characteristic function for the sum of independent random variables is given by the product of the individual characteristic functions}.  
Given the $\Nband$ random variables $Y_i$ and their probability  
distributions ${\cal P}_i$ (Eq. \ref{eq:approx}), 
we calculate the distribution of the random variable $z \equiv  
\sum_i Y_i$, which will represent the goodness of fit,  as follows:
for each $Y_i$ we can compute the corresponding characteristic function $\Phi_i(k)$. Then using the property cited above, we can construct the characteristic function $\Phi_z(k)$ of the variable $z$ by  $\Phi_z(k) = \Phi_1(k)...\Phi_{{\Nband}}(k)$. The probability distribution function of  $z$, ${\cal F}(z)$, is then given by the inverse Fourier transform of $\Phi_z(k)$.
This approach is particularly straightforward in our case 
because the probability function given in Eq. (\ref{eq:approx})  
is just a $\chi^2$ law with $\nu_i$ DOF,  
whose characteristic function $\Phi_i$ has an analytic form.  
Multiplication of the individual characteristic 
functions thus gives an analytical expression whose inverse  
Fourier transform is itself a $\chi^2$ distribution  
in $z$, with ${\bf \nu} = \sum \nu_i$ DOF: 
\begin{eqnarray} 
{\cal F}(z) = z^{{\bf \nu}/2}\; e^{-z/2}\\ 
\nonumber \\ 
{\rm with} \;\;\; z = \sum_i Y_i \;\;\;\nonumber {\rm and}   
     \;\;\; {\bf \nu} = \sum \nu_i \nonumber  
\end{eqnarray} 
The variable $gof=z$ is thus (another) $\chi^2$--distributed 
quantity that provides a useful GOF statistic.

\section{``The good, the bad and the GOF'' or  
Are CMB fluctuations consistent with a Gaussian  distribution?} 
 
\subsection{Application} 
 
     In this section we apply each of the above GOF statistics 
to the CMB data set shown in Figure~\ref{fig_plot};  
note that this does note 
include the most recent BOOMERanG, MAXIMA and DASI results. Adding these  new
data will essentially results in reducing the ``$\chi^2$'' distributed $gof$
  values without changing drastically the results presented in this section. 
Our overall approach is as described in Le~Dour et al. 2000 (hereafter paper 3) and Douspis et al. 2001a, where 
we used the likelihood approximation given in paper 1 
to find the best model. 
We consider three combinations of data:  
Data set 1 contains   
all points ({\bf ALL})\footnote{Actually, we noticed that 
our approximation fails to recover the MC simulations for upper 
limits. For this reason we do not include them in our analysis};  
set 2 consists of all the data minus the Python 5 results (Coble et al. 1999)
 ({\bf ALL-5}); 
and set 3 combines just COBE (Tegmark \& Hamilton  1997),
 MAXIMA (Hanany et al. 2000) and BOOMERanG (de Bernardis et al. 2000) 
({\bf CMB}). 
\begin{figure} 
\begin{center} 
\resizebox{\hsize}{!}{\includegraphics[angle=0,totalheight=6cm, 
        width=6cm]{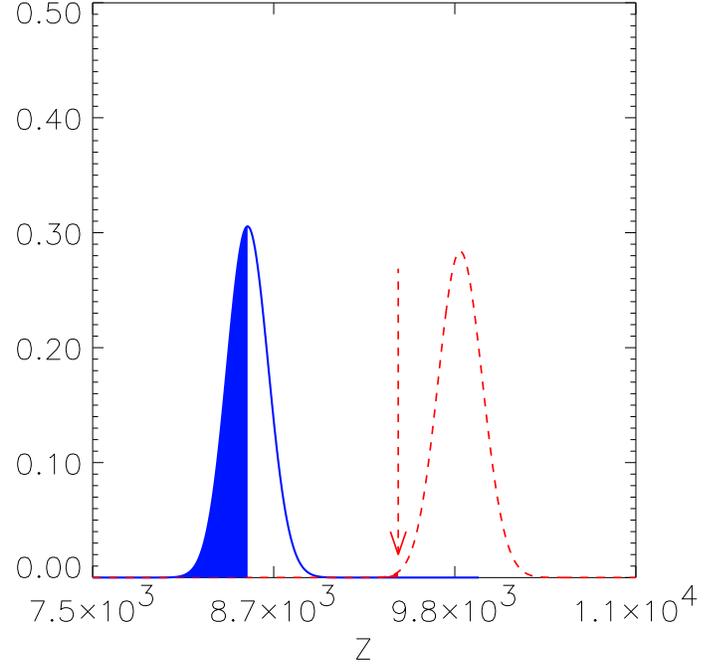}} 
\end{center} 
\caption{\label{f:resfc}Results of the GOF for each subset given by 
our CF test. The values of the GOF with the \cf technique are given 
by the  blue shaded part for ALL subset and the red arrow for the 
ALL-5  subset.} 
\end{figure} 
The best model for each data set will be  
referred to as $\rm BM_{ALL}$, $\rm BM_{ALL-5}$,$\rm BM_{CMB}$. 
A summary of the various GOF statistics for these models 
is given in Table 1; the lines are labelled by  
{\em GC} for ``generalised $\chi^2$'', {\em CF} for 
``characteristic functions'', and ``$\chi^2$'' for the 
classic $\chi^2$ of Eq.~ (\ref{eq:classicchi2})\footnote{We noticed that  
$\sigma_i$ is given different definitions in the literature. 
When considering the evaluation of the GOF using each definition, 
we found that the value of the GOF is quite sensitive to the definition  
of $\sigma_i$. We consider in this paper the technique giving the best  
value of the GOF}. 
 
\begin{table}[h] 
\label{t:table1} 
\begin{tabular}{|c||c|c|c|} 
\hline 
&ALL&ALL-5& CMB \\ 
\hline 
GC & 0.02\% & 8.6\%  & 60.0\% \\ 
CF &  0.3\% & 51.0\% & 63.0\% \\ 
$\chi^2$ & 0.004\% & 1.1\% & 55.0\% \\ 
\hline 
\end{tabular}      
\caption[]{Values of the GOF of the ``best models'' for\\ 
each subset of the actual data. GC is for ``generalised $\chi^2$'', CF for the ``characteristic functions'' technique and $\chi^2$ for the classic $\chi^2$.} 
\end{table} 
 

For GC technique, the $gof$ is directly equivalent to the absolute  
value of a $\chi^2$. To convert this into percentage, we need to know the DOF.
The latter is given in our case by the number of  
experiments taken into account in each set minus  the number of free  
cosmological parameters.  
 
For the CF technique, the percentage given in table 1 is obtained by  
integrating the probability distribution function $f$ of $Z_Y$ from infinity 
to 
$Z_{obs}=\sum_i^{N_{set}} Z_i^2$ where $N_{set}$ is the number 
 of experiments in each set.

  
The figure \ref{f:resfc} summarises the results given by our CF test on both
 data sets 1 and 2. The  line gives the function to 
integrate and the shaded part is the integrated  part corresponding to 
the  numbers given in Table \ref{t:table1}. The solid (blue) line and  
shaded part correspond to data set 1, and the dashed (red) line and  
arrow to data set 2.

\subsection{Discussion}

The first remark to be made based on Table 1 is that the complete 
data set (set 1) is {\em inconsistent} with a Gaussian sky fluctuations, 
according to all three techniques; the GC method, for example, 
excludes this hypothesis at more than 99.99 \%. 
This means in particular that it is not appropriate to search   
cosmological constraints, because the whole class of models 
considered is ruled out.  This could be due to several 
effects, in particular the fact that we do not include 
calibration uncertainties in our analysis. 
 
The situation is different if we remove Python 5 (set 2) from 
the analysis.  In this case,  our two evaluations of the GOF  
(GC and CF) both accept the hypothesis of Gaussian sky fluctuations. 
In contrast, the classic (but inappropriate) $\chi^2$ statistic  
marginally excludes such  hypothesis.  Figure 3 illustrates the difference  
between our GC method and the classic  $\chi^2$, data point by 
data point (for a subset of data set 1). Triangles show individual 
 $\chi^2$ values, while  
boxes correspond to the $\nu_i$ defined in section 3.  
We see that the classic $\chi^2$ overpenalizes the fit for 
outliers, a conclusion already noted in paper 2. 
 
Finally, we can see that all three methods accept the Gaussian 
hypothesis as a good representation to the COBE, MAXIMA and  
BOOMERanG data (set 3).  

\section{Conclusion} 
 
We have discussed three different ways of estimating the GOF 
to CMB band--powers.  A GOF statistic is a key element   
of any parameter estimation study, and a good fit must 
be insured before considering parameter constraints.  
The classic $\chi^2$ GOF statistic is not rigorously applicable 
to power spectrum data, because power estimates are not  
Gaussian distributed quantities.  We propose instead  
two alternative GOF statistics based on an approximation 
to the distribution of power estimators. This approximation 
was motivated by the same kind of arguments presented in  
paper 1 for the likelihood function.  The distribution 
of a power estimator is a different quantity than the likelihood 
function used to define the estimator. We tested the 
approximation presented here against Monte Carlos simulations 
of CMB observations and found that it  
reproduced well the distribution of the maximum likelihood 
band--power estimator.   
 
     We then constructed two different GOF statistics, whose 
distributions were found using the approximate  
power estimator distribution.  With the same, rather 
minimal information required to build the likelihood 
approximation (paper 1), we are now also able to develop 
a GOF statistic to test the quality of the 
maximum likelihood model to a set of band--power data, 
thereby allowing a complete statistical analysis of 
anisotropy data from diverse observations. 
The method is limited by the fact that we are unable 
to account for correlations between band--powers; this, 
however, is not a serious restriction, as these correlations 
are usually rather unimportant for the final results 
based on current data sets. 
 
     In applying this approach to a set of band--power data  of Figure 
\ref{fig_plot} we found that the ``best model'' obtained 
is in fact a bad fit. In other words, the data  
are unlikely to have been drawn from a Gaussian distribution 
represented by such a model.  The fit becomes acceptable  
if we exclude the Python 5 points from the analysis,  
according to our GOF statistics.  This is most likely 
due to the fact that we do not account for calibration errors, 
and so the bad fit probably just indicates that the adopted 
calibration is incorrect.  It is interesting to note that, 
even with Python 5 removed, the classic $\chi^2$ still marginally rejects 
the best fit.  We traced this behaviour to the fact that this 
method over weights the importance of ``outliers''. 
 
     The important cosmological conclusion is that this CMB data set 
(excluding Python 5, due to our inability to account for  
calibration errors) is consistent with Gaussian sky fluctuations 
drawn from the best--fit inflationary model.  

A final remark concerns the possibility offered by the development of an
approximated distribution function of the estimators. In the application
of current Monte Carlo methods for $C_\ell$'s extraction 
(eg. Szapudi et al. 2000, MASTER: Hivon et al. 2001), the estimator 
distribution is a natural output. The likelihood function needed in parameter
estimations is however unknown.
The present study suggests that we could   reconstruct the likelihood function directly {\it from} the estimator distribution. 
 The two parameters ($\nu$ and $\beta$) can be fitted on the 
estimator distribution and then used in the approximated likelihood function
of Bartlett et al. 2001. Consequently one is then able to reconstruct all the 
likelihood function and to perform a proper parameter estimation.

\begin{acknowledgements} 
M. D. would like to thank Nabila Aghanim for usefull comments and corrections.
\end{acknowledgements}


\begin{thebibliography}{} 
 
\bibitem[Balbi~{\it et~al.}{~2000}]{balbi}
Balbi, A. {\it et al.}~2000, ApJ, 545, L1
\bibitem{} Bartlett J.G., Blanchard A., Le Dour M., Douspis M. 
        \& Barbosa D. 1998a, in: Fundamental 
        Parameters in Cosmology (Moriond Proceedings),  
        Eds. J. Tr\^an Thanh V\^an et al. (Editions Fronti\`eres: 
        Paris, France), astro--ph/9804158 
\bibitem{} Bartlett J.G., Blanchard A., Douspis M. \& Le Dour M. 1998b, 
        to be published in: Evolution of Large--scale Structure: 
        from Recombination to Garching (Munich, Germany), 
        astro--ph/9810318 
\bibitem{} Bartlett J.G., Blanchard A., Douspis M. \& Le Dour M. 1998c, 
        to be published in: The CMB and the Planck Mission 
        (Santander, Spain), astro--ph/9810316 
\bibitem{} Bartlett J.G., Douspis M., Blanchard A. \& Le Dour M.
  2000, Astronomy and Astrophysics Supplement, v.146, p.507-517, 2000
(BDBL)
\bibitem{} Bond J.R., Jaffe A.H. \& Knox L., 2000, ApJ 533, 19  
\bibitem{} Bond J.R. \& Jaffe A.H., 1998,  
        In: Philosophical Transactions of the Royal Society  
        of London A, "Discussion Meeting on Large Scale 
        Structure in the Universe," Royal Society, London, March 1998, 
        astro--ph/9809043 
\bibitem{} Borrill J., 1999a, In: Maiani L., et al. (eds.) 3K Cosmology, 
        AIP Conf. Proc. 476, 277, astro--ph/9903204    
\bibitem{} Borrill J., 1999b, In: Proceedings of the 5th European  
        SGI/Cray MPP Workshop, astro--ph/9911389 
\bibitem{} Clapp, A.\ C.\ et al.\ 1994, ApJ Lett., 433, L57 
\bibitem{}Coble, K.\ et al.\ 1999, ApJ Lett., 519, L5 
\bibitem[de~Bernardis~{\it et~al.}{~2000}]{boom} de Bernardis, P.\ et al.\ 2000, Nature, 404, 955
\bibitem{} Dodelson S. \& Knox L. 2000, Phys.Rev.Lett. 84 (2000) 3523
\bibitem{} Douspis M., Bartlett, J.G., Blanchard A. \& Le Dour M.,
	Astronomy and Astrophysics, v.368, p.1-14, 2001a
\bibitem{} Douspis M.,  Blanchard A., 
	Sadat R.,  Bartlett, J.G.,\& Le Dour M.,  A\&A in press, 
	astro-ph/0105129, 2001b
\bibitem{} Efstathiou, G., Bridle, S.\ L., Lasenby, A.\ N., Hobson, 
	M.\ P., \& Ellis, R.\ S.\ 1999	, MNRAS, 303, L47 
\bibitem[Hanany~{\it et~al.}{~2000}]{maxima} Hanany, S.\ et al.\ 2000, ApJ Lett., 545, L5 
\bibitem{} Hancock S., Rocha G., Lasenby A.N. \& Gutierrez C.M. 1998,
        MNRAS 294, L1
\bibitem{} Hivon E., Gorski K.M., Netterfield C.B., Crill B.P., Prunet S.,  Hansen F., astro-ph/0105302, 2001
\bibitem{} Knox L. \& Page L., 2000, Phys.Rev.Lett. 85 (2000) 1366
\bibitem{} Lahav, O.\ \& Bridle, S.\ L.\ 1999, Evolution of Large Scale 
	Structure : From Recombination to Garching, 190 
\bibitem[Lange~{\it et~al.}{~2000}]{lange}
Lange, A.E.~{\it et~al.}~2000, Phys.Rev.D, 63, 42001
\bibitem{} Lasenby A.N., Bridle S.L. \& Hobson M.P. 1999, 
        to be published in: The CMB and the Planck Mission
        (Santander, Spain), astro--ph/9901303
\bibitem{} Le Dour, M., Douspis, M., Bartlett, 
	J.\ G., \& Blanchard, A.\ 2000, A\&A, 364, 369 
\bibitem[Lineweaver~{\it et~al.}{~1997}]{lineweaver} Lineweaver C., Barbosa D., Blanchard A. \& Bartlett J.G.
        1997, A\&A 322, 365
\bibitem{} Lineweaver C.H. \& Barbosa D. 1998a, A\&A 329, 799
\bibitem{} Lineweaver C.H. \& Barbosa D. 1998b, ApJ 496, 624
\bibitem{} Lineweaver C.H. 1998, ApJ 505, 69
\bibitem{} Netterfield, C.\ B., Devlin, M.\ J., Jarolik, N., Page, L., 
	\& Wollack, E.\ J.\ 1997, ApJ, 474, 47 
\bibitem[Netterfield~{\it et~al.}{~2002}]{boom2}Netterfield, C.~B.~{\it et~al.},~2002, ApJ, 571, 604
\bibitem[Pryke~{\it et~al.}{~2002}]{pryke}
Pryke, C.~{\it et~al.}~2002, ApJ, 568, 46
\bibitem[Rubino-Martin~{\it et~al.}{~2002}]{rubino}
Rubino-Martin, J. A.~{\it et~al.}~2002, {\tt astro-ph/0205367}
\bibitem{} Szapudi I., Prunet S., Pogosyan D, Szalay A.S., Bond J.R., astro--ph/0010256
\bibitem[Sievers~{\it et~al.}{~2002}]{sievers}
Sievers J. L.~{\it et~al.}~2002, ApJ, submitted, {\tt astro-ph/0205387}
\bibitem{} Tegmark M. \& Hamilton A. 1997, astro--ph/9702019
\bibitem{} Tegmark, M.\ \& Zaldarriaga, M.\ 2000a, ApJ, 544, 30 
\bibitem{} Tegmark M. \& Zaldarriaga M. 2000b,	Phys. Rev. Lett., 85, 2240 

\bibitem[Wang~{\it et~al.}{~2002}]{wang}
Wang, X., Tegmark, M., \& Zaldarriaga, M.~2002, Phys.Rev.D, 65, 123001
\bibitem{} Webster, A.\ M., Bridle, S.\ L., Hobson, M.\ P., Lasenby,
	 A.\ N., Lahav, O., \& Rocha, G.\ 1998, ApJ Lett., 509, L65 

\end{thebibliography}
\end{document}